\documentclass[reprint,amsmath,amssymb,aps,prb,superscriptaddress,showpacs,floatfix]{revtex4-2}
\usepackage{graphicx}
\usepackage[colorlinks=true,linkcolor=blue,citecolor=blue,urlcolor=blue]{hyperref}
\usepackage{dcolumn}
\usepackage{multirow}
\usepackage{bm}
\usepackage{mathtools}
\usepackage{tensor}
\usepackage[version=4]{mhchem}
\usepackage{mathrsfs}
\usepackage{slashed}
\usepackage{feynmp}
\usepackage{graphicx}
\usepackage{graphics}
\usepackage{placeins}
\usepackage{amssymb}
\usepackage{float}
\usepackage{mathtools}%可用于下面格式的排版，在符号左边添加角标%$\prescript{n}{m}{H}_i^j<L$
\usepackage{tensor}%主要用于张量的排版，\indices,\tensor两个命令
\usepackage{mhchem}%主要用于化学式的排版，\ce命令eg:\ce{H2O}
\usepackage{mathrsfs}
\usepackage{slashed}%量子场论中的字母上划斜线
\usepackage{verbatim}%李代数常用字体包
\usepackage{stackrel}%使用stackrel的包

\usepackage{feynmp}

\usepackage{amsmath,amssymb}
\allowdisplaybreaks
\newcommand\lowint%定义上下积分
{
\mspace{2mu}\underline{\vphantom{\int}\mspace{10mu}}\mspace{-9mu}\int
}

\newcommand\diracr[1]{|{#1}\rangle}%dirac右矢量
\newcommand\diracl[1]{\langle{#1}|}%dirac左矢量
\newcommand{\myhslashslash}{%
  \raisebox{.78ex}{%
    \scalebox{.7}{%
      \rotatebox[origin=l]{15}{$-$}%
    }%
  }%
}
\newcommand{\myhslash}[1]{%
  {%
   \vphantom{b}%
   \ooalign{\kern-0.12em\smash{\myhslashslash}\hidewidth\cr$#1$\cr}%
   \kern 0em
  }%
}

\usepackage{amsmath,amssymb}
\usepackage[normalem]{ulem}

\renewcommand{\sout}[1]{}
\renewcommand{\revised}[1]{#1}%

\begin{document}
\title{Regularized universal topological markers for Dirac systems}
\author{Yulin Qin}
\affiliation{Department of Physics, Fudan University, Shanghai 200433, China}
\affiliation{Department of Physics, School of Science, Westlake University, Hangzhou 310024, Zhejiang, China}
\author{Chang-An Li}
\affiliation{Hefei National Laboratary, Hefei 230088, China}
\affiliation{School of Emerging Technology, University of Science and Technology of China, Hefei 230026, China}
\affiliation{Institute for Theoretical Physics and Astrophysics, University of Würzburg, 97074 Würzburg, Germany}
\author{Jian Li}
\affiliation{Department of Physics, School of Science, Westlake University, Hangzhou 310024, Zhejiang, China}
\date{\today}
\begin{abstract}
\revised{Topological} markers provide an efficient and powerful characterization of topological features of many systems, especially when the translation symmetry is broken. Recently, a universal topological marker applicable in different symmetry classes of topological systems is proposed. However, it suffers from irregular behaviors at the boundary and its connection to other topological indexes remains elusive.  In this work, we construct regularized universal topological markers that apply to Dirac systems by utilizing position operators that are compatible with periodic boundary conditions. The regularized markers eliminate the obstructive boundary irregularities successfully and give rise to the desired global topological invariants such as the Chern number consistently when integrated over all the lattice sites. Furthermore, the regularized form allows us to establish an explicit connection between the markers and some other known topological indices in two dimensions. For instance, it turns out to be equivalent to the Bott index in classes A, D, and C and equivalent to the spin Chern number in classes DIII and AII. We further examine the utility and stability of this marker in disordered scenarios. We find that its variance shows peaks at the phase boundaries, which promotes it as a useful indicator for detecting disorder-induced topological phase transitions.    
\end{abstract}
\maketitle

\section{\label{sec:level1}Introduction}
Topological invariants serve as the most essential and fundamental characterizations of topological phases of matter in different symmetry classes \cite{Altland1997PhysicalReviewBp1142,Schnyder2008PRBp195125,Schnyder2009p10,Kitaev2009p22,Ryu2010NJoPp65010,Hasan2010RMPp3045,Qi2011RMPp1057,Chiu2016p35005p35005,Klitzing1980PRLp494,Laughlin1981PRB,Thouless1982PRLp405,Haldane1988,Su1980PRBp2099,Kitaev2001p131p131,Kane2005Prlp226801,Kane2005Prlp146802,Fu2006PRB,Bernevig2006sp1757,Bernevig2006PRLp106802}. In general, topological invariants are defined based on band theory and can be obtained by homotopy maps from the Brillouin zone (BZ) to a target manifold,  which relies on the translation symmetry of the interested systems. Topological properties can survive when the translation symmetry is broken, for instance, in the disordered system, while the calculation of topological invariants in real space is not straightfoward. Substantial progress has been made in the real space representation of the topological invariants. For example, the real space representation of the Chern number can be derived from the perspective of noncommutative geometry  \cite{Bellissard1994p5373p5451,Prodan2010NJoP,Prodan2010PRLp115501,Prodan2011,Xue2012PRBp155445,Bourne2018JoPAMaTp235202,Prodan2013JoPAMaT} and from the Bott index formula \cite{Exel1991JoFAp364,Hastings2010Jompp,Hastings2011AoPp1699,Loring2011}. There are also real space representation of the winding number for chiral symmetric systems \cite{MondragonShem2014p46802p46802,Lin2021p224208p224208} and the real space markers \cite{Bianco2011,Gebert2020PRAp63606,dOrnellas2022PRBp155124,Hannukainen2022PRLp277601,Sousa2023PRBp205133,Melo2023PRBp195151,Molignini2023SPCp59,Bau2024PRBp14206,Bau2024PRBp54203,Cerjan2024PRLp73803,Velury2021PRBp24205,Velury2025PRBp094204}. A natural question arises: Can these invariants be represented uniformly for \textit{any} symmetry class in \textit{any} spatial dimension?\par

Recently,  a topological invariant in momentum space called the wrapping number has been proposed along this direction  \cite{Gersdorff2021PRBp245146}. This progress further sheds light on its real-space representations. Subsequently, universal topological markers have been further constructed \cite{Chen2023PRBp45111}. However, 
there are two main issues with these universal topological markers. First, they do not respect translation symmetry even in clean systems. These markers are invalid near the boundaries of the lattice system, showing boundary irregularities, since the real space position operators employed therein are not periodic. Second, the desired global properties are not encoded in these markers. In other words, averaging these markers over all sites does not yield the desired global topological invariant, making them inapplicable when considering the influence of disorder on global properties. Although one ad hoc remedy for this issue by taking the modulus of markers has been proposed \cite{Oliveira2024PRBp94202}, a more rigorous approach to regularize these markers is still missing. \par

In this work, we present a regularized universal topological  marker that overcomes the two difficulties aforementioned and exhibits more rich physical consequences. Based on the degree of the orientation vector that describes Dirac systems, we construct a universal topological  marker by introducing position operators that are consistent with periodic boundary conditions. It thus shows translation symmetry globally and avoids the undesired boundary irregularities. The corresponding topological invariant, such as the Chern number, can further be obtained by summing the markers over all lattice sites.  Remarkably, based on this formula, we can address the direct connections between the universal topological markers and other topological invariants in two dimensions (2D). We find that this formula is equivalent to the Bott index for symmetry classes A, D, and C, and to the spin Chern number for classes DIII and AII. In addition, since the global properties are actually encoded in the alternative marker, its local fluctuations provide essential information about the disordered topological systems. Its variance peaks at the phase boundaries, which can be used as an indicator for disorder-induced topological phase transitions. 

The rest of the paper is organized as follows. In Sec. \ref{sec:level2}, we review the definition of the wrapping number for topological insulators and superconductors described by the Dirac model, and then present our regularized universal topological markers. In Sec. \ref{sec:level3}, we demonstrate the connection between the our formula and the Bott index formula in 2D. In Sec. \ref{sec:level4}, we provide numerical results by applying our formula to lattice systems so as to consider the influence of disorder and study the fluctuations of these markers. Finally, we present our conclusions in Sec. \ref{sec:level5}.

\section{\label{sec:level2}Regularized universal topological markers from the wrapping number}
The Dirac model describing topological insulators and superconductors in $D$  dimensions can be expressed in momentum space as \cite{Schnyder2008PRBp195125,Ryu2010NJoPp65010,Chiu2016p35005p35005}
\begin{equation}
H_0(\mathbf{k})=\mathbf{d}(\mathbf{k})\cdot\mathbf{\Gamma}=d(\bf k)\mathbf{n}(\mathbf{k})\cdot\mathbf{\Gamma},
\end{equation}
where $\mathbf{\Gamma}=(\Gamma_0,\Gamma_1,\cdots,\Gamma_D)$ is a vector composed of a subset of the Dirac matrices $\{\Gamma_0,\Gamma_1,\cdots,\Gamma_{2s}\}$ that satisfy the anticommutation relations $\{\Gamma_i,\Gamma_j\}=2\delta_{i,j}$ and $s=\lfloor\frac{D+1}{2}\rfloor$ or $s=\lfloor\frac{D+3}{2}\rfloor$, depending on the symmetry class \cite{Ryu2010NJoPp65010}; $\mathbf{d}(\mathbf{k})=(d_0(\mathbf{k}),d_1(\mathbf{k}),\cdots,d_D(\mathbf{k}))$ is a vector in $D+1$ dimensions that characterizes the momentum dependence of the model and $d(\bf k)=|\mathbf{d}(\mathbf{k})|$. Importantly, $\mathbf{n}(\mathbf{k})=\mathbf{d}(\mathbf{k})/|\mathbf{d}(\mathbf{k})|$ is the orientation vector of $\mathbf{d}(\mathbf{k})$, which is a map from the torus $T^D$ to the sphere $S^D$. According to the Hopf degree theorem \cite{milnor1997topology}, the homotopy classification of such maps is determined by the wrapping number or equivalently the degree of the map $\mathbf{n}(\mathbf{k})$, which is defined as
\begin{equation}
\mathrm{deg}[\mathbf{n}]=\int_{\text{BZ}}\frac{d^{D}\mathbf{k}}{V_{D}}\epsilon_{i_{0}\cdots i_{D}}n^{i_{0}}\partial_{k_1}n^{i_{1}}\partial_{k_2}n^{i_{2}}\cdots\partial_{k_D}n^{i_{D}},
\end{equation} 
where $V_D$ is the volume of the unit sphere $S^D$ in $D$ dimensions. It turns out that this wrapping number captures the topological invariants for various symmetry classes and spatial dimensions for topological insulators and superconductors, thus serving as a unified topological invariant \cite{Gersdorff2021PRBp245146}. Considering the flattened Hamiltonian $H_{\mathbf{k}}=\mathbf{n}(\mathbf{k})\cdot\mathbf{\Gamma}$, the wrapping number can be further expressed as \cite{Chen2023PRBp45111}
\begin{equation}
    \text{deg}[\mathbf{n}]=\int_{\text{BZ}} \frac{d^{D}\mathbf{k}}{2^{s}cV_{D}} \text{Tr}[WH_{\mathbf{k}}\partial_{k_{1}}H_{\mathbf{k}}\partial_{k_{2}}H_{\mathbf{k}}\cdots \partial_{k_{D}}H_{\mathbf{k}}],
\end{equation}
where $2^sc=\text{Tr}[\Gamma_{D+1}\Gamma_{D+2}\cdots\Gamma_{2s}\Gamma_0\Gamma_1\cdots\Gamma_{D}]$, and $c\in\{1,i,-i,-1\}$ depends on the ordering of these matrices and the value of $s$. The matrix $W$ is the  product of remaining Dirac matrices defined as $W\equiv\Gamma_{D+1}\Gamma_{D+2}\cdots\Gamma_{2s}$. 

In the following, we demonstrate how to construct a well-defined topological marker from this wrapping number. Without loss of generality, we consider a quadratic Hamiltonian for a free fermionic system with translation symmetry, which can be expressed in either the real space or the momentum space as
\begin{align}
H=\sum_{\bm{r},\bm{r}^\prime}H_{\bm{r},\bm{r}^\prime}\diracr{\bm{r}}\diracl{\bm{r}^\prime}=\sum_\mathbf{k}H_\mathbf{k}\diracr{\mathbf{k}}\diracl{\mathbf{k}},
\end{align}
where we have absorbed all inner degree of freedom in the position basis $\diracr{\bm{r}}$ or the momentum basis $\diracr{\mathbf{k}}$ for brevity. Defining the momentum translation operator along the $i$th direction (i.e., the real space position operator) as
\begin{align}
    \hat{X}_i=\sum_{\mathbf{k}}\diracr{\mathbf{k}-\delta \mathbf{k}_i}\diracl{\mathbf{k}}=\sum_{\bm{r}}e^{-i\delta\mathbf{k}_i\cdot\bm{r}}\diracr{\bm{r}}\diracl{\bm{r}},
\end{align}
with $\delta \mathbf{k}_i=\frac{2\pi}{L_i}\mathbf{e}_i$, where $L_i$ is the total number of unit cells along the $i$-th direction (\revised{we choose $L_i=L$ in the following context}) and $\mathbf{e_i}$ is the unit vector. We then obtain a useful relation
\begin{equation}
\hat{X}_i H\hat{X}_i^\dagger=\sum_\mathbf{k}H_{\mathbf{k}+\delta \mathbf{k}_i}\diracr{\mathbf{k}}\diracl{\mathbf{k}},
\end{equation}
which helps us to transfer the intergral in the wrapping number as  
%\begin{samepage}
\revised{
\begin{align}
\notag\int& d^{D}\mathbf{k}\text{Tr}\Big[WH_{\mathbf{k}}\partial_{k_{1}}H_{\mathbf{k}}\partial_{k_{2}}H_{\mathbf{k}}\cdots\partial_{k_{D}}H_{\mathbf{k}}\Big]\\
\notag=&\frac{1}{2^D}\sum_\mathbf{k}\text{Tr}\Big[ WH_\mathbf{k}\prod_{i=1}^{D}(H_{\mathbf{k}+\delta \mathbf{k}_i}-H_{\mathbf{k}-\delta \mathbf{k}_i})\Big]+O(\frac{1}{L^2})\\
=&\frac{1}{2^D}\text{Tr}\Big[WH\prod_{i=1}^{D}(X_iHX_i^\dagger-X_i^\dagger HX_i)\Big]\hspace{0.5cm} (L\rightarrow \infty).
\end{align}
}
%\end{samepage}
Note that we have written all the operators in real space, using $W$ to represent $I_N\otimes W$ for brevity, where $I_N$ is the $N$ dimensional identity matrix. Consequently, the wraping number is writen as
\begin{equation}
\text{deg}[\mathbf{n}]=\frac{1}{2^{s}cV_{D}2^D}\text{Tr}\Big[WH\prod_{i=1}^{D}(X_iHX_i^\dagger-X_i^\dagger HX_i)\Big],
\end{equation}
where $X_i=e^{-i\frac{2\pi}{L_i}\hat{x}_i}$ is real space representation of the momentum translation operator along the $i$-th direction. 

Finally, we arrive at the following universal topological operator
\begin{equation}
    \hat{C}=\frac{N}{2^{s}cV_{D}2^D}WH\prod_{i=1}^{D}(X_iHX_i^\dagger-X_i^\dagger HX_i),\label{equ:real-topo}
\end{equation}
where the factor $N$ is defined as \revised{$N=\prod_{i=1}^D L_i=L^D$}, which is the total number of unit cells.
We denote its local value
\begin{equation}
\label{equ:improved-marker}C(\bm{r})=\sum_\ell\diracl{\bm{r},\ell} \hat{C} \diracr{\bm{r},\ell}
\end{equation}
as a universal topological marker, where  $\ell$ indicates the inner degrees of freedom at site $\bm{r}$.  From this definition, the wrapping number can be expressed as
\begin{equation}
    \label{equ:global-invariant2}\text{deg}[\mathbf{n}]=\frac{\text{Tr}[\hat{C}]}{N}\equiv \bar{C}.
\end{equation}
 Equations (\ref{equ:real-topo})-(\ref{equ:global-invariant2}) present the main results of our work.

 \begin{figure*}
    \centering
    \includegraphics[width=1.0\linewidth]{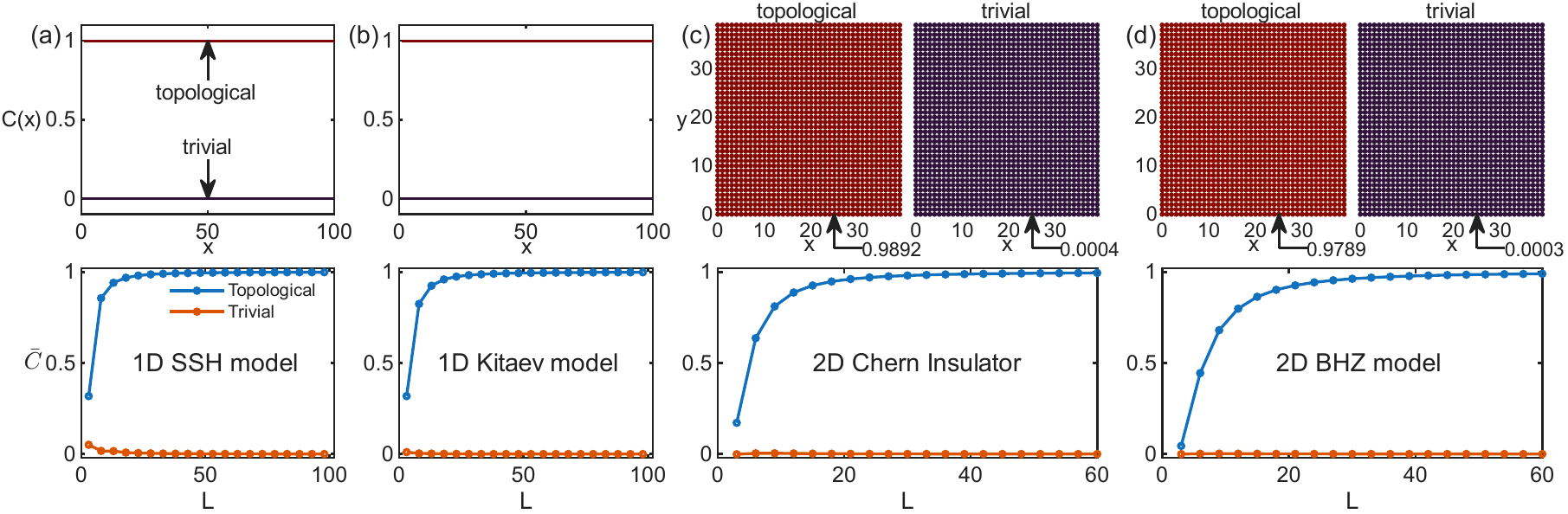}
    \caption{\revised{Distribution of the} regularized universal topological markers \revised{and the scaling behavior of these markers} for various models: (a) for 1D SSH model, (b) for 1D Kitaev chain, (c) for 2D Chern insulator, and (d) for 2D BHZ model. Top panel: Distribution of the regularized universal markers; in (a) and (b), we use a lattice of 100 unit cells; in (c) and (d), we use a lattice of $40\times 40$ unit cells. Bottom panel: Scaling behavior of the averaged markers where we choose $L_x=L$ and $L_x=L_y=L$ for one-dimensional and two-dimensional models respectively. The parameters used in each model are detailed in Appendix \ref{sec:app3}. In all cases, the topological markers distribute uniformly and converge to the expected topological invariant representing the corresponding phase.}
    \label{fig:local_markers}
\end{figure*}

This universal topological marker $C(\bm{r})$ demonstrates several remarkable properties. First, it is fully compatible with periodic boundary conditions. The position operator $X_i$  in Eq.~(\ref{equ:real-topo}) is well-defined for lattice Hamiltonian. It maintains translation symmetry for clean systems. If we translate the whole system along the $i$th direction, a global phase is introduced in operator $X_i$. This global phase will be canceled by $X_i^\dagger$, thus the whole expression remains invariant. Therefore, the markers $C(\bm{r})$ are uniformly distributed at site $\bm{r}$, even when $\bm{r}$ is close to the boundaries. This means that $C(\bm{r})$ does not suffer from boundary irregularities for finite-size lattices as seen in \cite{Chen2023PRBp45111}.  Explicitly, we apply our formula to prototypical topological models, i.e., the Su-Schrieffer-Heeger (SSH) model \cite{Su1980PRBp2099} and Kitaev chain model \cite{Kitaev2001p131p131} in 1D, and Chern insulator and Benervig-Hughes-Zhang (BHZ) model \cite{Bernevig2006PRLp106802} in 2D. We present the main results in Fig. \ref{fig:local_markers} and leave the details of these models to Appendix~\ref{sec:app3}.  It is found that the topological markers are uniformly quantized at $C(\bm{r})$ =1 for topological nontrivial phases and $C(\bm{r})$=0 for trivial phases, which provide proper characterization for the corresponding models. Second, the global topological invariant is entirely encoded in these markers. This feature, being more relevant when disorder is taken into account, enables us to derive the precise global topological invariant by averaging the markers over all the lattice sites. As such, the universal topological marker $C(\bm{r})$ permits an explicit connection with other real space representation of topological invariants such as the Bott index formula for 2D system, as shown in the following Sec.~\ref{sec:level3}.  Third, the fluctuations of these markers could serve as an indicator for topological phase transitions induced by disorder whereas the average value of these markers represents the global topological invariant. We will elaborate on this point in detail in Sec.~\ref{sec:level4}. 

\section{\label{sec:level3}Connection with the Bott index formula in two dimensions}
The Bott index formula is a powerful tool to calculate the Chern number \cite{Hastings2010Jompp} and the spin Chern number \cite{Huang2018PRBp125130}. In this section, we explicitly demonstrate the connection between our topological markers and the Bott index in 2D. 

\subsection{\label{sec:level3-1}Symmetry classes A, D, and C}
For symmetry classes A, D, and C in 2D, we consider the minima Dirac model whose flattened Hamiltonian in monmentum space can be expressed as $H(\mathbf{k})=\mathbf{n}(\mathbf{k})\cdot\bm{\sigma}$, where $\bm{\sigma}=(\sigma_x,\sigma_y,\sigma_z)$ is the Pauli matrix vector. In these cases, $W$ is the identity matrix, $2^sc=\text{Tr}[\sigma_x\sigma_y\sigma_z]=2i$. The corresponding wrapping number is 
\begin{align}
    \notag\text{deg}[\mathbf{n}]&=\frac{1}{8\pi i}\text{Tr}[H(XPX^\dagger-X^\dagger PX)(YPY^\dagger-Y^\dagger PY)]\\
    \notag&=\frac{1}{2\pi i}\text{Tr}[\frac{1}{2}(QXPYQX^{\dagger}PY^{\dagger}-QXPY^{\dagger}QX^{\dagger}PY)\\
    \label{equ:Bott-tuidao1}&\phantom{=\frac{1}{2\pi i}}-\frac{1}{2}(PXPYPX^{\dagger}PY^{\dagger}-PXPY^{\dagger}PX^{\dagger}PY)],
\end{align}
%where we have used the fact that $H=Q-P=1-2P$ with $P(Q)$ being the projection operator onto the occupied(un-occupied) states, $X,Y,X^\dagger,Y^\dagger$ are position operators commuting with each other, as well as the cyclical invariance under trace operation.
where we have used the relation $H=Q-P=1-2P$ with $P(Q)$ being the projection operator onto the occupied(un-occupied) states, the commutativity of the position operators $X,Y,X^\dagger,Y^\dagger$, and the cyclic property of the trace operation.\par
It is well-known\cite{Hastings2010Jompp,Hastings2011AoPp1699,Toniolo2022LiMPp126} that the norms $||[X,P]||$ and $||[Y,P]||$ are of order $O(1/L)$ for a local, bounded Hamiltonian with a finite gap, therefore
\begin{align}
\notag &\text{Tr}[\frac{1}{2}(QXPYQX^{\dagger}PY^{\dagger}-QXPY^{\dagger}QX^{\dagger}PY)]\\
\notag =&\text{Tr}[\frac{1}{2}([X,P]P[P,Y][X^\dagger,P]P[P,Y^\dagger]\\
\notag &\phantom{\text{Tr}[}-[X,P]P[P,Y^\dagger][X^\dagger,P]P[P,Y])]\\
\label{equ:vinishing-term}\sim &\revised{O(\frac{1}{L^2})}\xrightarrow{L\rightarrow \infty}0.
\end{align}
Here, we have used $QAP = (1-P)AP = [A,P]P$ and $PAQ = PA(1-P) = P[P,A]$ for arbitrary $A$. It is worth noting that in the presence of certain additional symmetry, for instance the particle-hole symmetry $S$ which relates $P$ and $Q$ by $SPS^{-1}=Q$ and $SQS^{-1}=P$, Eq. (\ref{equ:vinishing-term}) is valid regardless of the lattice size. To show this, notice that $S$ is the antiunitatry operator, thus we have 
\begin{align}
\notag  &\text{Tr}[QXPYQX^{\dagger}PY^{\dagger}]\\
\notag =&\text{Tr}[SQXPYQX^{\dagger}PY^{\dagger}S^{-1}]^*\\
\notag =&\text{Tr}[PX^\dagger QY^\dagger PXQY]^*\\
\notag =&\text{Tr}[(PX^\dagger QY^\dagger PXQY)^\dagger]\\
=&\text{Tr}[QXPY^{\dagger}QX^{\dagger}PY].
\end{align}
In general, we conclude
\begin{align}
    \notag\text{deg}[\mathbf{n}]&=\frac{1}{2\pi i}\text{Tr}[\frac{1}{2}(PXPY^{\dagger}PX^{\dagger}PY-PXPYPX^{\dagger}PY^{\dagger})]\\
    \label{equ:Bott}&=\frac{1}{2\pi i}\text{Tr}[\frac{1}{2}(U_YU_XU_Y^\dagger U_X^\dagger-U_XU_YU_X^\dagger U_Y^\dagger)],
\end{align}
\revised{where we have defined nonsingular matrices $U_{X}=V^\dagger X V$ and $U_{Y}=V^\dagger Y V$ with $V=(\diracr{\psi^{\text{occ}}_{1}},\diracr{\psi^{\text{occ}}_{2}},\cdots)$ being the matrix composed of all eigenvectors for the occupied states}. Further denoting $U=U_YU_XU_Y^\dagger U_X^\dagger$, we obtain \cite{Hastings2011AoPp1699,Toniolo2018PRBp235425}
\begin{align}
    \notag\text{deg}[\mathbf{n}]&=\frac{1}{2\pi i}\text{Tr}[\frac{1}{2}(U-U^\dagger)]=\frac{1}{2\pi }\text{Im}(\text{Tr}[U])\\
    \notag &= \frac{1}{2\pi }\text{Im}(\text{Tr}[\text{log}(U)])+\revised{O(\frac{1}{L^2})}\\
   \label{equ:Bott1} &= \frac{1}{2\pi}\text{Im}(\text{Tr}[\text{log}(U)])\hspace{0.5cm} (L\rightarrow \infty).
\end{align}
The right side of this equation is nothing but the Bott index formula in 2D. Finally, we arrive at the important relation
\begin{equation}
  \label{equ:barc-bott}\bar{C}=\frac{1}{2\pi }\text{Im}(\text{Tr}[\text{log}(U)]),
\end{equation}
which connects our topological markers to the Bott index. 

\begin{figure}
\centering
\includegraphics[width=\linewidth,height=4cm]{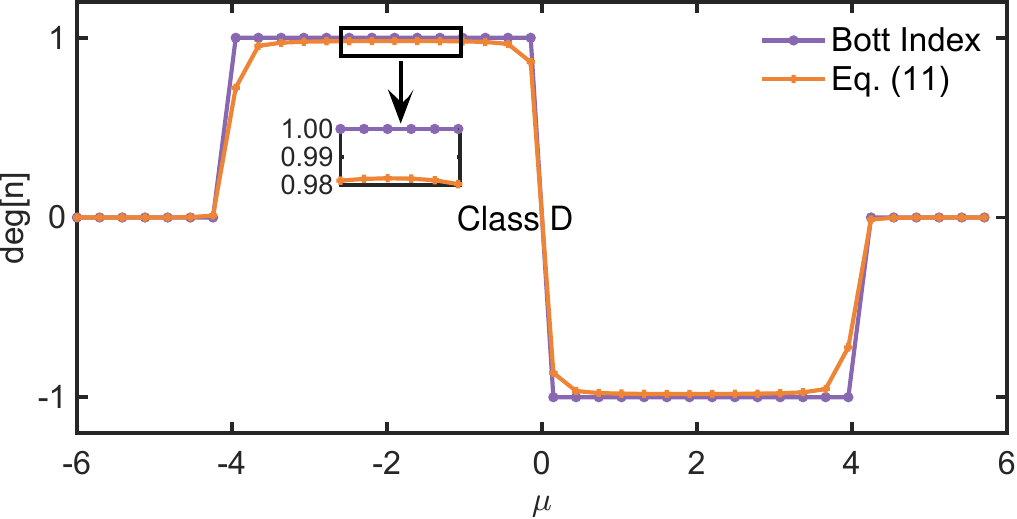}   
\caption{\label{fig:bott-D}Comparison for topological markers and the Bott index formula to verify Eq.~\ref{equ:barc-bott}. We use the two-dimensional system of symmetry classes D in this numerical simulation. The parameters are $t=-1,\Delta=0.5$. The inset provides a close-up for partial detail. The sign convention for $\text{deg}[\mathbf{n}]$ is chosen to ensure consistency with the Chern number in this figure. }
\end{figure}

To demonstrate this connection more transparently, we consider an explicit D class Dirac Hamiltonian given in momentum space by \cite{Chiu2016p35005p35005}
\begin{align}
\notag H_D(\mathbf{k})&=(2t\cos k_{x}+2t\cos k_{y}-\mu)\sigma_{z}\\
&+2\Delta\sin k_{x}\sigma_{x}+2\Delta\sin k_{y}\sigma_{y},
\end{align}
where the particle-hole operator can be expressed as $\sigma_x K$ with $K$ being the complex conjugation and the product of remaining Dirac matrices is $W=\sigma_0 $.  Here, $t$  is the hopping integral, $\mu $ is the chemical potential, and $\Delta$  is the pairing amplitude. We employ a lattice of $40\times 40$ unit cells for numerical simulation and the results are presented in Fig.~\ref{fig:bott-D}. It is clear that our topological markers are consistent with the Bott index formula \cite{Hastings2010Jompp}, capturing the topological phases and phase transitions successfully. The slight deviation of our results from the Bott index  could arise from finite size effect. Therefore, the averaged topological markers provide a reliable  approximation to the Bott index formula.\par

\subsection{\label{sec:level3-2}Symmetry classes DIII and AII}
For symmetry classes DIII and AII, we employ the strategy of dimensional reduction to show the connections directly  \cite{Ryu2010NJoPp65010}. In 2D, three Dirac matrices 
$\{\gamma_1,\gamma_2,\gamma_3\}$ out of the five 
$\{\gamma_1,\gamma_2,\gamma_3,\gamma_4,\gamma_5\}$ 
are employed to construct the minimal Dirac Hamiltonian as
\begin{equation}
H_{DIII(AII)}(\mathbf{k})=n_1(\mathbf{k})\gamma_1+n_2(\mathbf{k})\gamma_2+n_3(\mathbf{k})\gamma_3.
\end{equation}
In this case, $W$ is the product of two Dirac matrices that commutes with $H$. Thus we can further define 
$W=\gamma_4\gamma_5\equiv iS_z$,
where $S_z$ is Hermitian and traceless. The $S_z$ squares to identity, and we call it the generalized spin polarization operator. Since $2^sc=\text{Tr}[\gamma_4\gamma_5\gamma_1\gamma_2\gamma_3]=(2i)^2$ in this case, and note that $I_N\otimes W$ also commutes with $X,Y,X^\dagger$ and $Y^\dagger$, we arrive at 
 \begin{align}
     \notag\text{deg}[\mathbf{n}]&=\frac{1}{2i}\frac{1}{2\pi i}\text{Tr}[\frac{1}{2}W(PXPY^{\dagger}PX^{\dagger}PY\\
     \notag&\phantom{=\frac{1}{2i}\frac{1}{2\pi i}\text{Tr}[\frac{1}{2}W(}-PXPYPX^{\dagger}PY^{\dagger})]\\
    \notag&=\frac{1}{2}\frac{1}{2\pi i}\text{Tr}[\frac{1}{2}S_z(PXPY^{\dagger}PX^{\dagger}PY\\
    \label{equ:2D-DIII-AII}&\phantom{=\frac{1}{2}\cdot\frac{1}{2\pi i}\text{Tr}[\frac{1}{2}S_z(}-PXPYPX^{\dagger}PY^{\dagger})],
 \end{align}
where we have shortened $I_N\otimes W (I_N\otimes S_z)$ as $W (S_z)$ for brevity. On the other hand, $S_z=S_+-S_-$, where $S_+(S_-)$ is the projection onto the $+1(-1)$ eigenspace of the operator $S_z$. \revised{In addition, the eigenvectors for the occupied states can be divided into two sectors $\{\diracr{\psi^{\text{occ}}_{1,+}},\diracr{\psi^{\text{occ}}_{2,+}},\cdots\},\{\diracr{\psi^{\text{occ}}_{1,-}},\diracr{\psi^{\text{occ}}_{2,-}},\cdots\}$ which correspond to the eigenvectors of $S_z$ with $\pm1$ eigenvalues respectively. Denoting $V_{\pm}=(\diracr{\psi^{\text{occ}}_{1,\pm}},\diracr{\psi^{\text{occ}}_{2,\pm}},\cdots)$ to be the matrix composed of all the eigenvectors for the occupied states from these two sectors accordingly}, we have
 \begin{align}
   \notag&\text{Tr}[S_zPXPY^{\dagger}PX^{\dagger}PY]\\
   \notag=&\text{Tr}[(S_+-S_-)PXPY^{\dagger}PX^{\dagger}PY]\\ 
   \notag=&\text{Tr}[P_+XP_+Y^{\dagger}P_+X^{\dagger}P_+Y-P_-XP_-Y^{\dagger}P_-X^{\dagger}P_-Y] \\
   =&\text{Tr}[U_{Y,+}U_{X,+}U_{Y,+}^\dagger U_{X,+}^\dagger -U_{Y,-}U_{X,-}U_{Y,-}^\dagger U_{X,-}^\dagger],
 \end{align}
where we have used $S_++S_-=I$ and  $S_\pm AS_\mp=0$ for any $A\in\{P,X,Y,X^\dagger,Y^\dagger\}$, which hold because these operators commute with $S_z$. \revised{Furthermore, we define the projected operators $P_{\pm}=S_{\pm}PS_{\pm}=V_{\pm} V_{\pm}^\dagger$ and nonsingular matrices $U_{X,\pm}=V_{\pm}^\dagger X V_{\pm}$ and $U_{Y,\pm}=V_{\pm}^\dagger Y V_{\pm}$.} A similar expression can be derived for the second term in equation (\ref{equ:2D-DIII-AII}). Further denoting $U_{\pm}=U_{Y,\pm}U_{X,\pm}U_{Y,\pm}^\dagger U_{X,\pm}^\dagger$, we obtain
 \begin{align}
     \notag\text{deg}[\mathbf{n}]&=\frac{1}{2}\frac{1}{2\pi i}\text{Tr}[\frac{1}{2}(U_+-U_-)-\frac{1}{2}(U_+^\dagger-U_-^\dagger)]\\
     \notag&=\frac{1}{2}\frac{1}{2\pi }\text{Im}(\text{Tr}[U_+]-\text{Tr}[U_-])\\
     \notag&=\frac{1}{2}\frac{1}{2\pi }\text{Im}(\text{Tr}[\log (U_+)]-\text{Tr}[\log (U_-)])+\revised{O(\frac{1}{L^2})}\\
     \label{equ:bott-spin-chern}&=\frac{1}{2}(C_+-C_-)\hspace{0.5cm}(L\rightarrow \infty),
 \end{align}
 which equates our topological marker to the Bott index formula for the spin Chern number with generalized spin polarization operator $S_z=-iW$ and $C_\pm=\frac{1}{2\pi }\text{Im}(\text{Tr}[\log(U_\pm)])$\cite{Huang2018PRBp125130,Huang2018PRLp126401}.\par

\section{\label{sec:level4}Disorder effect on the universal topological markers}
One of the most important advantage of topological markers lies in their applications in disordered systems when the translation symmetry is lost. Due to the remarkable properties mentioned above, our regularized topological markers not only give rise to the global topological invariants by integrating over all lattice sites but also provide signatures of topological phase transitions via their local fluctuations in disordered systems. In the following, we demonstrate the direct influence of disorder on topological markers, taking 1D BDI and D class topological models as concrete examples. 
\subsection{\label{sec:level4-1} Class BDI in one dimension}
\begin{figure}
    \centering
    \includegraphics[width=1.0\linewidth,height=10cm]{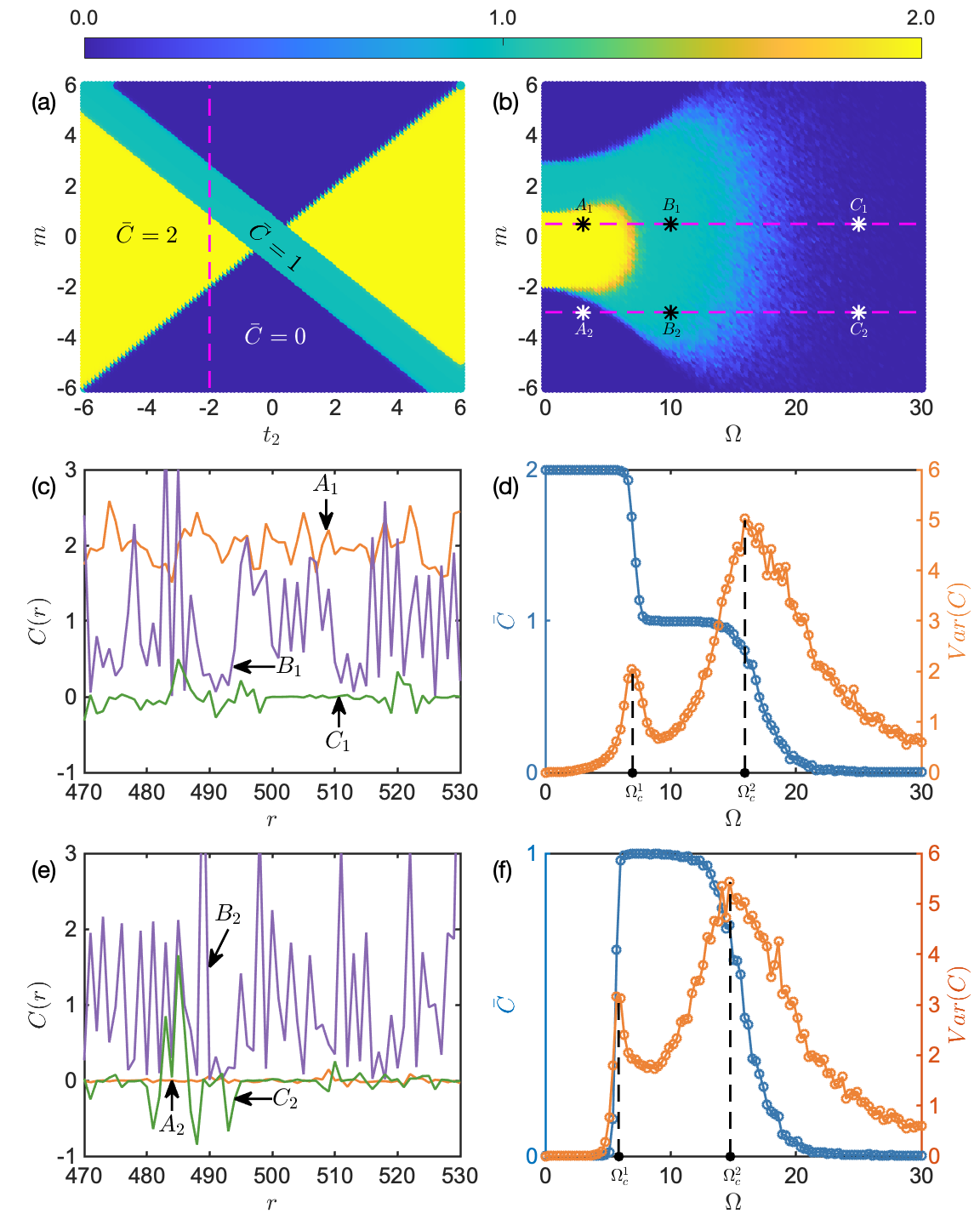}
    \caption{
    (a),(b) Phase diagrams of the extended SSH model: (a) Clean system as a function of $m$ and $t_2$ with $t_1=1.0$ and $\Omega=0$, the dashed purple line represent the case $t_2=-2.0$. (b) Disordered phase diagram as a function of $m$ and disorder strength $\Omega$ with $t_1=1.0$ and $t_2=-2.0$. We have averaged over ten disorder realizations. The two dashed purple lines indicate the cases $m=0.5$ and $m=-3.0$, respectively. $A_1,B_1,C_1$ are three points with specific disorder strengths on the line $m=0.5$. The corresponding values of markers at these representative points are presented in (c). Similarly, $A_2,B_2,C_2$ are three points for the case $m=-3$. (c), (e) \revised{Topological} markers $C(r)$(only 60 central sites are shown) at specific disorder strengths $\Omega=3,10$ and $25$ for a representative disorder realization for $m=0.5$ and $m=-3.0$, respectively. (d), (f) Global topological invariant $\bar{C}$ (blue curve, obtained by averaging all the markers) and corresponding variance $\mathrm{Var}(C)$ (orange curve) are plotted as functions of the disorder strength $\Omega$ for $m=0.5$ and $m=-3.0$, respectively. Each data point is averaged over 100 disorder realizations with system size of 1000 unit cells.
    }
    \label{fig:extended-ssh-disorder}
\end{figure}

For symmetry class BDI in 1D, we consider the extended SSH model. The disordered Hamiltonian with next-nearest couplings can be expressed as \cite{Song2014PRBp224203}
\begin{equation}
    \label{equ:extended-ssh-model2} H_1=\sum_i (t_{0,i}c_{i,A}^\dagger c_{i,B}+t_{1,i}c_{i+1,A}^\dagger c_{i,B}+t_{2,i}c_{i+2,A}^\dagger c_{i,B})+H.c.,
\end{equation}
 where the parameters are set to be $t_{0,i}=m+\Omega\omega_i$, $t_{1,i}=t_1+\frac{\Omega}{2}\omega_i$, and $t_{2,i}=t_2$. Here, $c^\dagger_{i,A(B)}$ and $c_{i,A(B)}$ are the creation and annihilation operators at site $i$ for the A(B) sublattice, respectively. The parameter $\Omega$ represents the disorder strength and $\omega_i$ is a random number distributes uniformly in $[-\frac{1}{2},\frac{1}{2}]$. For the clean system ($\Omega=0$), the phase diagram with respect to $m$ and $t_2$ by fixing $t_1=1.0$ is illustrated in Fig. \ref{fig:extended-ssh-disorder}(a), which is obtained by averaging all the markers with a lattice size of 500 unit cells. It describes the phase diagram of the extended SSH model properly \cite{Song2014PRBp224203}. 

The phase diagram with respect to $m$ and $\Omega$ \revised{at fixed parameters $t_1=1.0$ and $t_2=-2.0$} is illustrated in Fig. \ref{fig:extended-ssh-disorder}(b). \revised{We observe that the averaged value of these topological markers remains quantized in the yellow ($\bar{C}=2.0$) and green ($\bar{C}=1.0$) areas, demonstrating robustness against weak disorder. This agrees with earlier findings showing that the averaged marker is robust against weak disorder that alters the nonzero elements of the pristine Hamiltonian. Such robustness has been proven analytically via perturbation theory and discussed extensively for various symmetry classes in Ref.~\cite{Oliveira2024PRBp94202}. From this phase diagram, we also observe a smooth crossover between these quantized regions, which is consistent with previous studies \cite{Song2014PRBp224203,Oliveira2024PRBp94202}}.\par
For a detailed inspection on the influence of disorder on the global topological invariant and the phase transition induced by disorder, we consider two lines $m=0.5$ and $m=-3$, indicated in Fig. \ref{fig:extended-ssh-disorder}(b). The former one undergoes $\bar{C}=2\rightarrow \bar{C}=1\rightarrow \bar{C}=0$ if we gradually increase the disorder strength $\Omega$. In contrast, the latter one undergoes $\bar{C}=0\rightarrow \bar{C}=1\rightarrow \bar{C}=0$, which indicates a topological Anderson phase transition in the extended SSH model \cite{Li2009Prl,Song2014PRBp224203,LiCA20prl}.

In the presence of disorder, the markers $C(\bm{r})$ are no longer uniformly distributed at each lattice  site $\bm{r}$. Figures \ref{fig:extended-ssh-disorder}(c) and \ref{fig:extended-ssh-disorder}(e) present the values of markers for some representative points marked in phase diagram Fig. \ref{fig:extended-ssh-disorder}(b), which clearly show the existence of fluctuations of these markers in disordered systems. To investigate their local fluctuations, we  calculate the variance of these markers, namely,
\begin{equation}
    \text{Var}(C)=\frac{1}{N}\sum_{\bm{r}}(C(\bm{r})-\bar{C})^2. 
\end{equation}
Note that the system contains 1000 lattice sites in numerical calculations, and the  100 disorder realizations are averaged. The results are presented in Figs.~\ref{fig:extended-ssh-disorder}(d) and ~\ref{fig:extended-ssh-disorder}(f) for the two lines $m=0.5$ and $m=-3$, respectively. For the case $m=0.5$, it is clear that at weak disorder strength, the global topological invariant $\bar{C}$ remains robust against disorder, maintaining at the value $\bar{C}=2$. The corresponding variance of the markers increases with the disorder strength. At a critical value of $\Omega_{c}^1\revised{(=6.97\pm 0.20)}$, the variance reaches a local maximum. It then decreases within a small window around $\Omega_{c}^1$. Correspondingly, the global topological invariant gradually drops from $\bar{C}=2$ to $\bar{C}=1$, indicating a topological phase transition. Once the system enters the $\bar{C}=1$ phase,  the variance of the  markers increases until it reaches its next local maximum at $\Omega_{c}^2\revised{(=15.93\pm 0.30)}$, signaling another phase transition from $\bar{C}=1$ to $\bar{C}=0$ as increasing $\Omega$. For the case $m=-3$, similar results can be obtained. Thus, these  peaks of the variance can serve as indicator of phase transitions induced by disorder.

\subsection{\label{sec:level4-2} Class D in one dimension}
\begin{figure}
    \centering
    \includegraphics[width=1.0\linewidth,height=10cm]{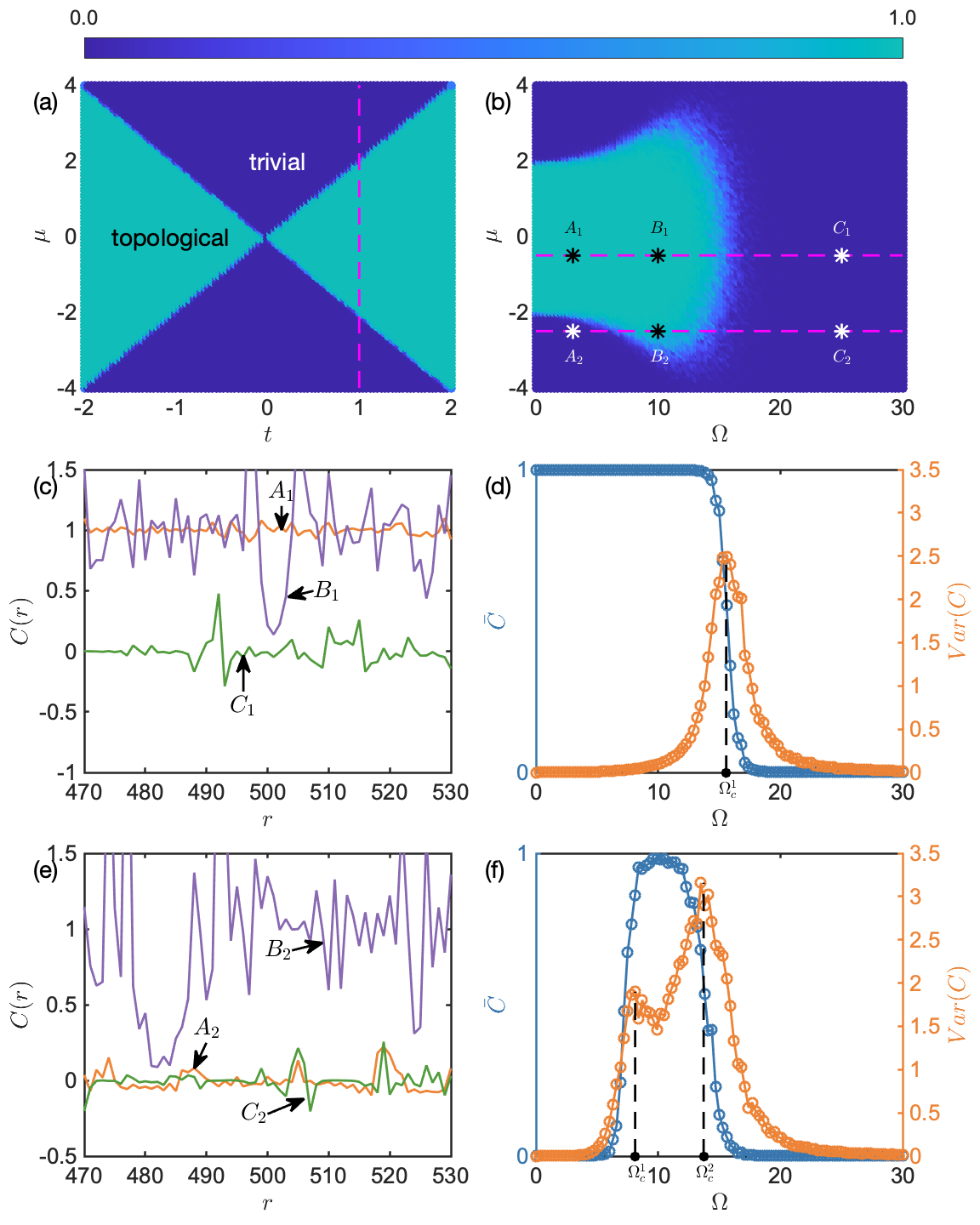}
    \caption{(a), (b)  Phase diagrams of one-dimensional D class Kitaev chain. (a) Clean system as a function of $\mu$ and $t$ with $\Delta=2.5$ and $\Omega=0$. The dashed purple line represent the case $t=1.0$. (b) Disordered phase diagram as a function of $\mu$ and disorder strength $\Omega$ with $t=1.0$ and $\Delta=2.5$. We have averaged ten disorder realizations. The two dashed purple lines indicate the cases $\mu=-0.5$ and $\mu=-2.5$, respectively. $A_1,B_1,C_1$ are three points with specific disorder strengths on the line $\mu=-0.5$, and the corresponding values of markers at these representative points are presented in (c). Similarly, $A_2,B_2,C_2$ are three points for the case $\mu=-2.5$. (c), (e) \revised{Topological} markers $C(r)$(only 60 central sites are shown) at specific disorder strengths $\Omega=3,10$ and $25$ for a representative disorder configuration for $\mu=-0.5$ and $\mu=-2.5$, respectively. (d), (f) Global topological invariant $\bar{C}$ (blue curve, obtained by averaging all markers) and variance $\mathrm{Var}(C)$ (orange curve) are plotted as functions of the disorder strength $\Omega$ for $\mu=-0.5$ and $\mu=-2.5$, respectively. Each data point is averaged over 100 disorder realizations with system size of 1000 unit cells.}
    \label{fig:D-1D-kitaev}
\end{figure}

For symmetry class D in 1D, we consider the Kitaev chain \cite{Kitaev2001p131p131} with random onsite chemical potentials. The Hamiltonian is 
\begin{align}
     \notag H_2=&\sum_i -\mu_i(c_i^\dagger c_i-\frac{1}{2})-t\sum_i(c_i^\dagger c_{i+1}+c_{i+1}^\dagger c_i)\\
    \label{equ:1D-D-realspace}&+\Delta\sum_i( c_{i+1}^\dagger c_i^\dagger+ c_{i} c_{i+1}),
\end{align}
where $\mu_i=\mu-\Omega\omega_i$, $\Omega$ represents the disorder strength and $\omega_i$ is a random number distributes uniformly in $[-\frac{1}{2},\frac{1}{2}]$. Here, $t$ and $\Delta$ are real parameters of this model. In the clean limit ($\Omega=0$), the system falls into the topological nontrivial phase when $|\mu|<2|t|$ and $\Delta\neq 0$. The phase diagram with respect to $\mu$ and $t$ by fixing $\Delta=2.5$ is illustrated in Fig. \ref{fig:D-1D-kitaev}(a). In the presence of disorder, the phase diagram with respect to $\mu$ and $\Omega$ is illustrated in Fig. \ref{fig:D-1D-kitaev}(b) for fixed $\Delta=2.5$ and $t=1.0$. \revised{Here, we again observe the robustness of the averaged topological markers against weak disorder, as well as a smooth crossover between the quantized regions\cite{Oliveira2024PRBp94202}}. The degree of Dirac Hamiltonian only yields a $Z$-valued invariant $\text{deg}(\mathbf{n})$, while the Kitaev chain falls into a $Z_2$ classification. This $Z_2$ invariant can be expressed as the wrapping number in the form $(-1)^{\text{deg}(\mathbf{n})}$ \cite{Gersdorff2021PRBp245146} or equivalently $\text{deg}(\mathbf{n})\hspace{0.4em}\text{mod}(2)$. To obtain the phase diagrams as shown in Figs.~\ref{fig:D-1D-kitaev}(a) and ~\ref{fig:D-1D-kitaev}(b), we have employed $|\text{deg}(\mathbf{n})|\hspace{0.4em}\text{mod}(2)$ as the $Z_2$ invariant by considering $-\text{deg}(\mathbf{n}) \equiv \text{deg}(\mathbf{n})\hspace{0.4em}\text{mod}(2)$. Similarly, we consider two lines $\mu=-0.5$ and $\mu=-2.5$, indicated by the dashed lines in Fig. \ref{fig:D-1D-kitaev}(b). The former one undergoes the transition $\bar{C}=1\rightarrow \bar{C}=0$ and the latter one undergoes the transition $\bar{C}=0\rightarrow \bar{C}=1\rightarrow \bar{C}=0$,  as the disorder strength $\Omega$ is gradually increased. The behavior of the markers, the global topological invariant, and the corresponding variance at specific disorder strengths $\Omega$ are shown in Figs. \ref{fig:D-1D-kitaev}(c)-\ref{fig:D-1D-kitaev}(f), which is much similar to the case of BDI class discussed before. Thus we can conclude that the local maxima of the variance of the markers can also serve as indicators of phase transitions induced by disorder in the 1D D class Kitaev chain.

\section{\label{sec:level5} Conclusion and discussion}
In summary, we have introduced a regularized universal topological marker applicable to any symmetry class and spatial dimension for topological insulators and superconductors described by the Dirac Hamiltonian, constructed from position operators that remain fully compatible with periodic boundary conditions. This formulation successfully removes boundary irregularities and, when summed over all lattice sites, reproduces the expected global topological invariants—such as the Chern number—in a consistent manner. Leveraging this property, we established explicit correspondences between our marker and several established topological indices in two dimensions, demonstrating its equivalence to the Bott index in symmetry classes A, D, and C, and to the spin Chern number in symmetry classes DIII and AII. Finally, we assessed the robustness of the marker in the presence of disorder and showed that its variance exhibits pronounced peaks at phase boundaries, highlighting its effectiveness as an indicator of disorder-driven topological phase transitions.

We note that the current version of universal topological markers are applicable only to topological insulators and superconductors that can be described by the Dirac model, at least at low energy. In these cases, the system’s flattened Hamiltonian in momentum space can be expressed as $H(\mathbf{k})=\mathbf{n}(\mathbf{k})\cdot \mathbf{\Gamma}$ where the central object in this framework is the degree of \(\mathbf{n}\), which is a map from \(T^D\) to \(S^D\) for a Dirac Hamiltonian in \(D\) dimensions. Note also that our regularized universal topological markers can be reduced to the one proposed by Chen \cite{Chen2023PRBp45111} in the thermodynamic limit (see Appendix~\ref{sec:app2}).
\revised{As for the numerical calculation of the topological markers, the most computationally intensive part is computing the projection operator $P$, which requires diagonalizing the real-space Hamiltonian $H_0$. For large system sizes, diagonalization becomes prohibitively slow and memory-intensive. The kernel polynomial method, which approximates the projection operator $P$ by expanding it in terms of Chebyshev polynomials of the sparse matrix $H_0$, turns out to be an efficient approach \cite{Weisse2006RMPp275,Varjas2020PRRp13229,Chen2025PRBp165304,Roy2026PRBp054201,Romeral2025PRBp134201}. This method makes it feasible to compute numerous topological markers across large systems.}
There are several \revised{directions} which call for further investigations in the future. \revised{First}, whether it is possible to uniformly represent all topological invariants for any symmetry class and spatial dimension beyond the Dirac models. One possible strategy is to represent the product of all the unused Dirac matrices—denoted by \(W\) in the main text—by using the local symmetry operators, since only the matrix \(W\) carries the information inherited from the Dirac matrices in Eq. (\ref{equ:real-topo}). For example, in 2D class A, the matrix \(W\) is simply the identity within the Dirac Hamiltonian system, conveying no additional information about the Dirac matrices. Thus, it may be natural to extend the scope of Eq. (\ref{equ:real-topo}) in this context to any system belonging to symmetry class A. Prodan et al. developed a real space formula for the Chern number for any system that closely resembles the formula derived from this strategy \cite{Prodan2010PRLp115501}. It remains unclear whether this strategy is valid when \(W\) is not the identity or for cases beyond symmetry class A, and this issue requires further investigation. \revised{Second, another important aspect of the topological operator involves non-local markers, which are defined as the off-diagonal elements of the topological operator  \cite{Molignini2023SPCp59,Chen2023PRBp45111}. It has been shown that the oscillation properties of these non-local markers provide information about the gap-closing points in the Brillouin zone, and their decay lengths can detect topological phase transitions \cite{Chen2023PRBp45111}. Thus, the behavior of these non-local markers provides a powerful tool for studying the process of topological phase transitions.} \revised{Third, in recent years, the quantum geometry becomes an important aspect of topological system. In 2D, as far as we known, the quantum metric markers also suffer from the boundary irregularities \cite{Bianco2011,Marrazzo2019PRLp166602,Marsal2024PRLp026002a,Romeral2025PRBp134201} for finite system and exploring proper formulism using methods developed here to resolve this problem will be the subject of our future work.}
\iffalse
It is known that the extent the Wannier functions of the system are bounded below by the integrated quantum metric\cite{Marzari1997PRBp12847,Marzari2012RMPp1419}.
\fi

\section{\label{sec:level6} Acknowledgments}
This work was supported by the National Natural Science Foundation of China under Grants No. 92265201 and No. 12574176, and the Innovation Program for Quantum Science and Technology under Project No. 2021ZD0302704. C.A.L. was financially supported by the start-up funding at HFNL (Grant No. QD2022600001) and  Würzburg-Dresden Cluster of Excellence ct.qmat, EXC2147, Project-Id No.390858490.

\appendix

\section{\label{sec:app3}\uppercase{Numerical results on the universal topological markers}}
To show the effectiveness of the regularized universal topological markers, we use the following four Dirac models as concrete examples in our numerical simulation since they cover all four kinds of topological invariants in the tenfold classification framework \cite{Altland1997PhysicalReviewBp1142,Schnyder2008PRBp195125,Schnyder2009p10,Kitaev2009p22,Ryu2010NJoPp65010,Hasan2010RMPp3045,Qi2011RMPp1057,Chiu2016p35005p35005,Chiu2016p35005p35005}. We list the Hamiltonian of these models in momentum space  to clearly show that they all belong to Dirac system.  The corresponding \revised{topological} markers are presented in Fig.~\ref{fig:local_markers} in the main text.
\begin{enumerate}
\item{ 
1D SSH model \cite{Su1980PRBp2099} (the winding number)
\begin{align}
H_0(k)=&(t_0+t_1\cos k)\sigma_{x}+t_1\sin k\sigma_{y}.
\end{align}
The time reversal operator is $\sigma_0K$, particle hole operator is $\sigma_z K$, and the chiral operator is $\sigma_z$.The product of all the remaining Dirac matrices is $W=\sigma_z$. This model Hamiltonian falls into the symmetry class BDI with a Z classification, where the associated topological invariant is called the winding number. We fix $t_1=1.0$ and consider two values of $t_0$ representing distinct phases: $t_0=0.5$ for the topological phase (winding number 1) and $t_0=1.5$ for the trivial phase.
}
\item{ 
1D Kitaev chain \cite{Kitaev2001p131p131}(Chern-Simons $Z_2$)
\begin{align}
H_0(k)&=(-2t\cos k-\mu)\sigma_{z}+2\Delta\sin k\sigma_{y}.
\end{align}
The particle hole operator is $\sigma_x K$. The product of all the remaining Dirac matrices is $W=\sigma_x$. This model Hamiltonian falls into the symmetry class D with a $Z_2$ classification, where the associated topological invariant is called Chern-Simons $Z_2$ \cite{Chiu2016p35005p35005}. We fix $t=1,\Delta=0.5$ and consider two values of $\mu$ representing distinct phases: $\mu=1.0$ for the topological phase and $\mu=3.0$ for the trivial phase.
}
\item{
2D Chern insulator (the Chern number)
\begin{align}
\notag H_0(\mathbf{k})=&(M+4B-2B\cos k_x-2B\cos k_y)\sigma_z\\
&+2A\sin k_x\sigma_x+2A\sin k_y\sigma_y.
\end{align}
The particle hole operator is $\sigma_x K$. The product of all the remaining Dirac matrices is $W=\sigma_0$. This model Hamiltonian falls into the symmetry class D with a $Z$ classification, where the associated topological invariant is called a Chern number \cite{Klitzing1980PRLp494,Thouless1982PRLp405,Haldane1988,Chiu2016p35005p35005}. We fix $A=1,B=1$ and consider two values of $M$ representing distinct phases: $M=-2$ for the topological phase (Chern number 1) and $M=2$ for the trivial phase.
}
\item{ 
2D BHZ model \cite{Bernevig2006PRLp106802,Bernevig2006sp1757}(Fu-Kane $Z_2$) 
\begin{align}
\notag H_0(\mathbf{k})=&2\Delta\sin k_{x}s_{x}\otimes \sigma_{z}+2\Delta\sin k_{y} s_{y}\otimes \sigma_0\\
+&\{2t(\cos k_{x}+\cos k_{y})-\mu\}s_{z}\otimes \sigma_0.
\end{align}
The time reversal operator is $s_0\otimes (-i\sigma_y) K$, the particle hole operator is $s_x\otimes \sigma_0 K$, and the chiral operator is $s_x\otimes \sigma_y$. The product of all the remaining Dirac matrices is $W=is_0\otimes \sigma_z$. This model Hamiltonian falls into symmetry class DIII with a $Z_2$ classification, where the associated topological invariant is called the Fu-Kane $Z_2$ invariant in literature\cite{Kane2005Prlp226801,Kane2005Prlp146802,Fu2006PRB,Bernevig2006PRLp106802,Bernevig2006sp1757,Chiu2016p35005p35005}. We fix $t=-1$, $\Delta=0.5$ and consider two values of $\mu$ representing distinct phases: $\mu=-3.0$ for the topological phase and $\mu=-5.0$ for the trivial phase.
}
\end{enumerate}

\section{\label{sec:app2}\uppercase{Reduced form of our regularized universal topological  markers}}
\begin{figure}
    \centering
    \includegraphics[width=1.0\linewidth]{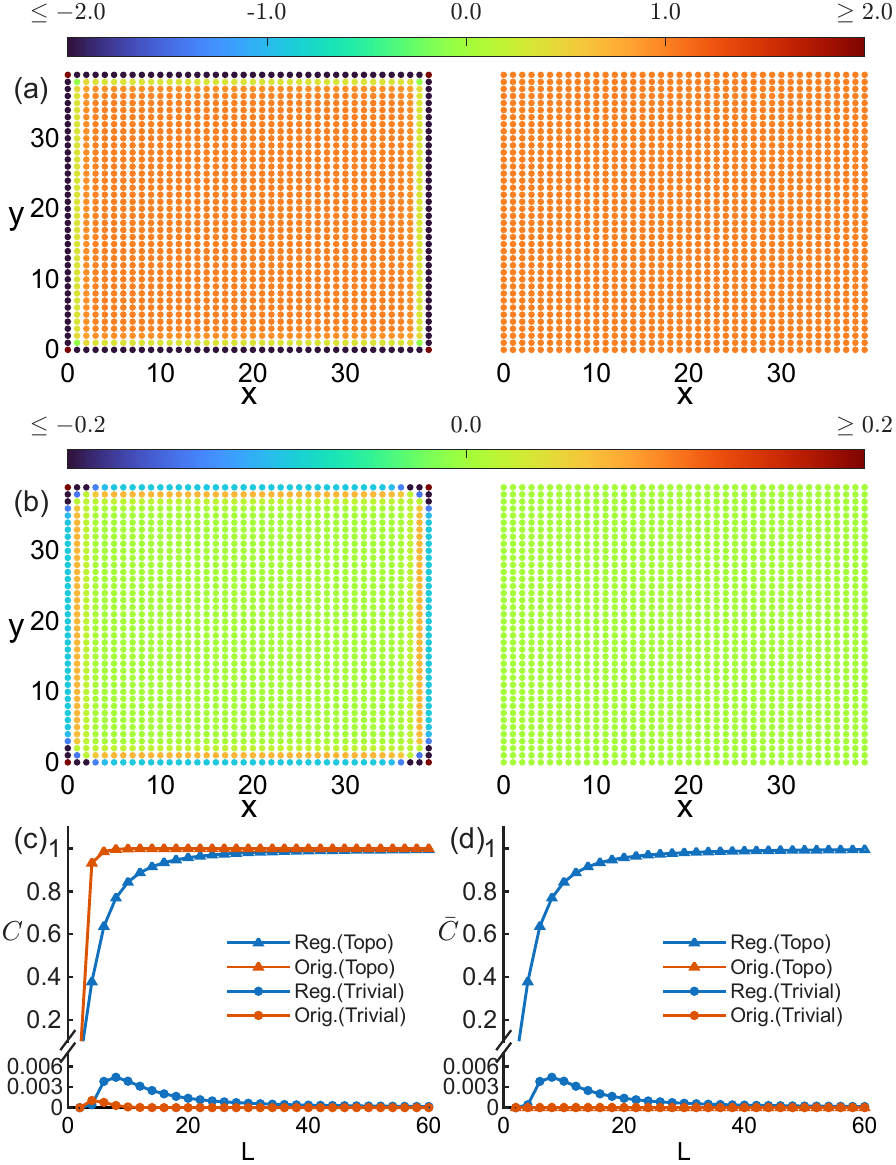}
    \caption{\revised{Comparison between the original topological markers proposed in \cite{Chen2023PRBp45111} and our regularized universal topological markers for the 2D Chern insulator. (a), (b) Spatial distribution of both marker types for the topological and trivial phases on a system size of $40\times 40$ unit cells. Left panels show original markers; right panels show regularized markers. (c) Center values of the original (Orig.) and regularized (Reg.) markers as functions of system size $L_x=L_y=L$ for topological (Topo) and trivial (Trivial) phase. (d) Averaged markers over all lattice sites versus system size for both phases. The original markers exhibit boundary irregularities and average to zero, whereas our regularized markers are boundary-free and correctly recover the expected Chern number upon spatial averaging.}}
    \label{fig:compare_markers}
\end{figure}

In this appendix, we show that our regularized universal topological markers can be reduced to the markers proposed in \cite{Chen2023PRBp45111} in the thermodynamic limit. First, the commutator  $[\hat{x}_i, H]$ is bounded regardless of the lattice size for a local Hamiltonian with a finite gap \cite{Hastings2010Jompp,Hastings2011AoPp1699,Toniolo2022LiMPp126}; in the thermodynamic limit, the relevant part is only the leading term
\begin{align}
   \label{equ:approx1}\frac{1}{2}(X_iHX_i^\dagger-X_i^\dagger HX_i)\sim-i\frac{2\pi}{L_i}[\hat{x}_i,H].
\end{align}
On the other hand
\begin{align}
   \label{equ:alter1} H[\hat{x}_i,H]&=-2(Q\hat{x}_iP+P\hat{x}_iQ),
\end{align}
Then, our universal topological operator is reduced to (we have used $H^2=I$)
\begin{widetext}
\begin{equation}
    \label{equ:topo-operator} C=N_D(I_N\otimes W)(Q\hat{x}_1P+P\hat{x}_1Q)(Q-P)(Q\hat{x}_2P+P\hat{x}_2Q)\cdots(Q-P)(Q\hat{x}_iP+P\hat{x}_iQ)\cdots (Q-P)(Q\hat{x}_DP+P\hat{x}_DQ)
\end{equation}
\end{widetext}
\noindent with $N_D=i^D2^{2D}\pi^D/2^scV_D$. By considering $QP=PQ=0$, only two terms survive in Eq. (\ref{equ:topo-operator}), which are precisely the alternating order terms between $P$ and $Q$, namely,
\begin{align}
    \notag C&=N_D(I_{N}\otimes W)[Q\hat{x}_{1}P\hat{x}_{2}Q\cdots \hat{x}_{D}O\\
    \label{equ:topo-operator1}&+(-1)^{D-1}P\hat{x}_{1}Q\hat{x}_{2}P\cdots \hat{x}_{D}\bar{O}],
\end{align}
where we have absorbed a possible minus sign into the coefficient c in $N_D$, and $\{O,\bar{O}\}=\{Q, P\}$ for the even D case and $\{O,\bar{O}\}=\{P,Q\}$ for the odd D case owing to the alternating ordering of projectors Q and P. Equation (\ref{equ:topo-operator1}) is just the topological operator defined in \cite{Chen2023PRBp45111}. \revised{To compare the original markers and our regularized markers clearly, we consider the two distinct phases in the 2D Chern insulator mentioned in Appendix~\ref{sec:app3} as an example. The numerical results are shown in Fig.~\ref{fig:compare_markers}. We can see that these two markers converge to the same value in the thermodynamic limit; however, for finite system sizes, the original markers suffer from boundary irregularities and always average to zero, whereas our regularized markers are free from these irregularities and correctly yield the expected Chern number when summed over all lattice sites which is an important feature that enables us to investigate the effects of disorder on global properties and to establish connections with other formulas, such as the Bott index, as discussed in the main text.}

\FloatBarrier

\end{document}